%% file: main.tex
\documentclass[sigconf]{acmart}
\AtBeginDocument{%
  }

\setcopyright{acmlicensed}
\copyrightyear{2026}
\acmYear{2026}
\acmDOI{XXXXXXX.XXXXXXX}
\acmConference[HRI '26]{ACM/IEEE International Conference on Human-Robot Interaction}{March 16--19,
  2026}{Edinburgh, Scotland, UK}

\usepackage{multirow}
\usepackage[table,xcdraw]{xcolor}
\usepackage{hhline}
\usepackage{graphicx}
\usepackage{url}
\usepackage{balance}
\usepackage{romannum}
\usepackage{float}
\usepackage{subcaption}  
\begin{document}

\title{LLM-Glasses: GenAI-Driven Glasses with Haptic Feedback for Navigation of Visually Impaired People}


\author{Issatay Tokmurziyev, Miguel Altamirano Cabrera, Muhammad Haris Khan, Yara Mahmoud, and Dzmitry Tsetserukou}
\affiliation{%
  \institution{Skolkovo Institute of Science and Technology (Skoltech)}
  \city{Moscow}
  \country{Russia}
}
\email{{issatay.tokmurziyev, m.altamirano, haris.khan, yara.mahmoud, d.tsetserukou}@skoltech.ru}






\renewcommand{\shortauthors}{Tokmurziyev et al.}

\begin{abstract}
LLM-Glasses is a wearable navigation system which assists visually impaired people by utilizing YOLO-World object detection, GPT-4o-based reasoning, and haptic feedback for real-time guidance. The device translates visual scene understanding into intuitive tactile feedback on the temples, allowing hands-free navigation. Three studies evaluate the system: recognition of 13 haptic patterns with an average recognition rate of 81.3\%, VICON-based guidance with predefined paths using haptic cues, and an LLM-guided scene evaluation with decision accuracies of 91.8\% without obstacles, 84.6\% with static obstacles, and 81.5\% with dynamic obstacles. These results show that LLM-Glasses can deliver reliable navigation support in controlled environments and motivate further work on responsiveness and deployment in more complex real-world scenarios.
\end{abstract}



\maketitle

\section{Introduction}

Navigating in dynamic environments is one of the major challenges for people with visual impairments, especially when it comes to real-time obstacle detection and avoidance. While GPS-based systems provide valuable guidance, their dependence on visual or auditory feedback often increases mental load and makes them less effective in crowded or unpredictable settings. Therefore, there is a need for intuitive, non-intrusive navigation tools \cite{tang}.
\begin{figure}[t]
    \centering
    \includegraphics[width=0.48\textwidth]{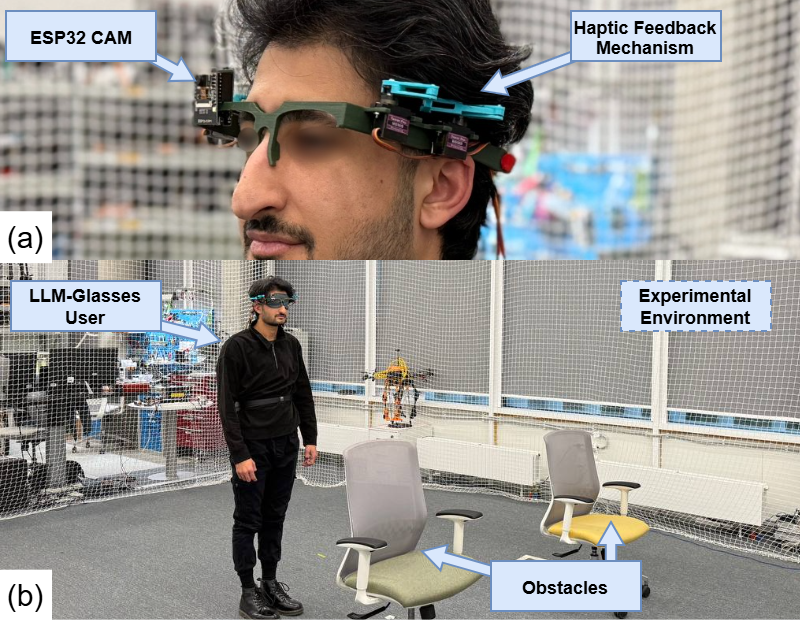} 
    \caption{(a) LLM-Glasses prototype, highlighting key components: ESP32 CAM and a haptic feedback mechanism with a five-bar linkage design; (b) an experimental setup with the user navigating through an obstacle course with real-time guidance provided by the system.}
    \Description{The figure contains two photographs side by side. Image (a) shows a close-up view of eyeglasses with an ESP32 CAM camera module mounted on the bridge and mechanical actuators attached to both temples using a five-bar linkage mechanism for delivering haptic feedback. Image (b) shows a person wearing the LLM-Glasses device while walking through an indoor space with chair obstacles, demonstrating real-world navigation assistance.}
    \label{fig:teaser}
    \vspace*{-0.4cm}
\end{figure}
Haptic feedback is a promising field for navigation support because it can apply spatial cues directly on the body while leaving vision and hearing available for other tasks. Prior work has explored a wide variety of haptic navigation devices, including augmented white canes, head- or waist-worn vibrotactile displays, and handheld or walker-mounted actuators that convey directional cues or obstacle proximity \cite{gallo,heuten,cabaret2024handle,navigation_haptics1,Kappers,iros2021running,lacote2024walker, dogsurf}. These systems have shown that haptic feedback can improve spatial perception, but many of them are bulky, require users to hold special hardware, or provide feedback that is not fully intuitive for continuous, hands-free use \cite{chinello,linkglide}. Recent findings on haptic and navigation technologies for blind and low-vision users show the trend toward multimodal solutions that combine tactile, auditory, and visual channels, as well as the need for lightweight, wearable interfaces that better integrate into everyday activities \cite{navigation_survey,haptic_review2024, gazegrasp}.

\begin{figure*}[t]
    \centering
    \includegraphics[width=0.85\textwidth]{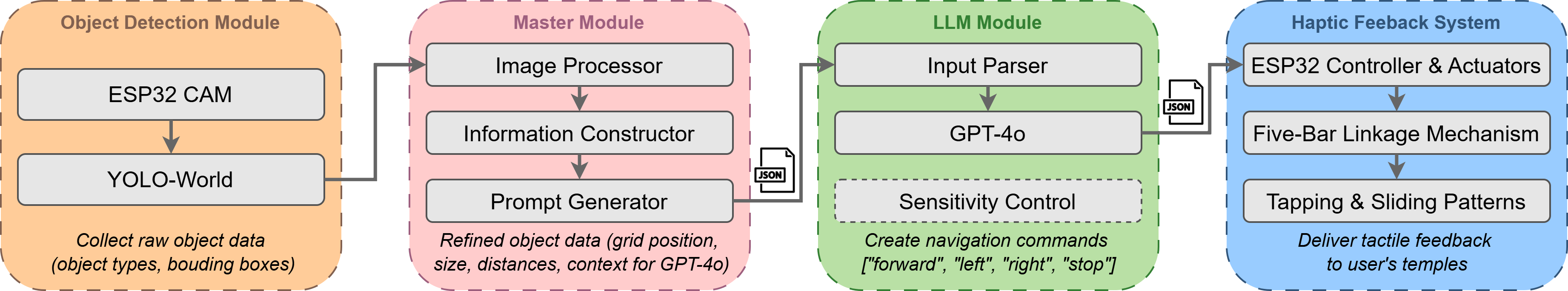} 
    \caption{System Architecture of the LLM-Glasses.}
    \Description{A system architecture diagram showing the data flow through four main modules. The Object Detection Module receives camera input from ESP32 CAM and uses YOLO-World to detect objects, outputting bounding boxes on a 2x3 spatial grid. The Master Module processes this data through image processing, multi-frame information construction, and prompt generation. The Language Module uses GPT-4o to convert structured prompts into navigation commands. Finally, the Haptic Feedback Module translates these commands into tactile patterns delivered through temple-mounted actuators controlled by ESP32. Arrows connect the modules showing the sequential processing pipeline from visual input to haptic output.}
    \label{fig:architecture}
\end{figure*}

Meanwhile, recent advances in computer vision and Large Language Models (LLMs) have created new opportunities for assistive navigation. Modern object detectors like YOLO-based frameworks can recognize many object classes in real time \cite{redmon2016lookonceunifiedrealtime}. Models like GPT-4o, BLIP, and LLaVA can translate complex scenes and generate personalized guidance for individual users \cite{li2022blipbootstrappinglanguageimagepretraining,ren2025touchedchatgptusingllm,xu2025llavacotletvisionlanguage,b2}. Several systems show how vision-language models can support blind users through dialogue interfaces and visual responses. For instance, VIAssist and ChatGPT interfaces describe scenes and suggest appropriate actions \cite{b11,b12, drone_companion}. However, these approaches have a common drawback: they rely primarily on speech or text-based feedback through smartphones. They may be useful for high-level navigation, but cannot provide the low-latency, embodied feedback needed for precise control when navigating in cluttered or dynamic environments. 

This paper introduces \emph{LLM-Glasses}, a lightweight wearable navigation system that combines YOLO-World object detection, GPT-4o-based reasoning, and temple-mounted haptic feedback to deliver real-time guidance without relying on audio output. ESP32 CAM captures the scene, detections are spatially organized and prioritized, and an LLM transforms this information into simple navigation commands that are rendered as intuitive tap and sliding tactile patterns on the user’s temples. The LLM-Glasses design utilizes two five-bar linkage mechanisms on the glasses frame to provide directional cues. To evaluate LLM-Glasses, three studies were conducted: a haptic pattern recognition experiment with 13 patterns, a VICON-based navigation task where participants followed predefined paths using only haptic feedback, and an LLM-guided video evaluation measuring decision accuracy in scenes with no, static, and dynamic obstacles. These experiments show that integrating haptic interfaces with CV and reasoning models can offer reliable navigation assistance in controlled environments and suggest potential advancements for more complex real-world scenarios.

\section{System Architecture}

\begin{figure*}[h]
    \centering
    \includegraphics[width=0.8\textwidth]{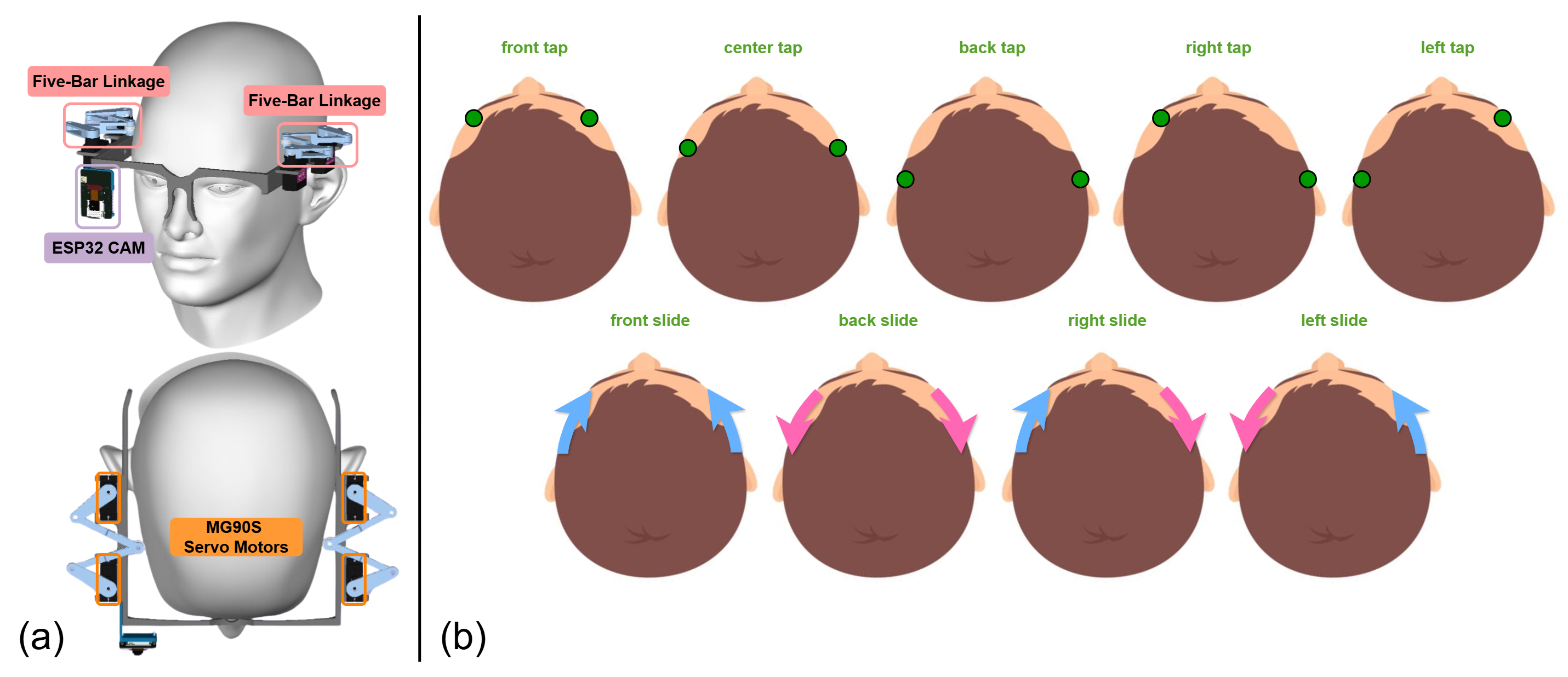} 
    \caption{(a) 3D model of the LLM-Glasses haptic navigation system; (b) the nine haptic feedback patterns used in the user study, illustrating tapping and sliding motions across different regions of the user’s temples. Each pattern provides distinct sensory input designed to aid in directional navigation.}
    \Description{Image (a) at the left, displays a 3D CAD rendering of the glasses frame showing the five-bar linkage mechanism mounted on the temple arms, with visible mechanical joints and actuator components in grayscale. Image (b) presents a diagram of nine distinct haptic patterns arranged in a grid, showing the temple area with arrows and dots indicating different tactile sensations. Five tap patterns are shown at different temple positions: front, center, back, left, and right. Four sliding patterns are illustrated with directional arrows: back-to-front, front-to-back, left-to-right, and right-to-left.}
    \label{fig:3dview}
\end{figure*}

LLM-Glasses is a lightweight wearable device that assists visually impaired users with real-time navigation by combining vision sensing, cloud-based language reasoning, and temple-mounted haptic feedback. The system processes camera images to detect and prioritize obstacles, converts this structured scene representation into compact prompts for a model, and renders the resulting navigation commands as intuitive tactile patterns on the user’s temples. The head temples were selected to apply haptic feedback to test the hypothesis regarding the ``hanger effect'' \cite{hanger}. Fig.~\ref{fig:architecture} illustrates the main modules: Object Detection, Master (Information Processing) Module, Language Module, and Haptic Feedback Module.

\paragraph{Object Detection Module}

The Object Detection Module serves as the system's sensory input, capturing real-time environmental data from ESP32 CAM. Frames are processed using the YOLO-World object detection model with bounding boxes and class labels for objects in the scene. Detected objects are classified and localized within a 2$\times$3 spatial grid in front of the user, and each object is assigned a priority based on its position and approximate distance so that closer obstacles in the lower central region are emphasized.

\paragraph{Master Module}

The Master Module acts as a bridge between object detection and language processing, converting raw detection data into organized and meaningful data. An image processing stage normalizes object positions and sizes across frames, while a multi-frame information constructor aggregates detections over short time windows to reduce momentary occlusions or detection errors. From this state, a prompt generator encodes object categories, grid locations, and proximity information into a concise textual description that conditions the language model, explicitly flagging nearby obstacles as high-priority hazards when they appear close to the user. The system ensures that the subsequent navigation commands prioritize obstacles that can be acted upon rather than distant objects.

\paragraph{Language Module}

The Language Module uses GPT-4o to interpret the structured scene description and produce simple navigation directives that can be rendered haptically. Given the current set of detected objects and their priorities, the model outputs high-level commands such as ``left'', ``right'', ``forward'', or ``stop'', optionally enriched with short justifications or contextual cues if additional detail is desired. The system incorporates advanced sensitivity controls, allowing users to adjust the level of detail in the feedback, from low (only critical hazard alerts) to high (more detailed obstacle reporting), enabling the same architecture to operate in both cluttered and relatively open spaces.

\paragraph{Haptic Feedback Module}

The Haptic Feedback Module converts navigation commands into tactile patterns delivered via temple-mounted actuators integrated into a lightweight glasses frame (see Fig. ~\ref{fig:3dview}a). The actuators are driven through a compact linkage mechanism and an ESP32-based controller, providing localized and sliding contact along the temples. The system uses a small set of taps and sliding patterns to encode direction and motion: taps signal discrete events such as confirmations, while sliding motions along the temples indicate continuous guidance cues (for example, sliding forward to move ahead or laterally to rotate). A brief calibration step adjusts intensity and contact position for each user, and the electronics are powered by separate battery packs for sensing/control and actuation, offering several hours of typical navigation use before recharging or replacement is required.

\section{Experimental Evaluation}

\input{confussion_matrix}
\paragraph{Haptic Pattern Recognition Study}
\label{section:user_study}

Temple-mounted tactile cues enable the system to guide head orientation changes during navigation. Using a five-bar linkage mechanism across a 7 cm range, the device delivers tap and sliding patterns that vary in force, position, direction, and speed. This study identifies patterns with the highest recognition rates for practical use.

We evaluated nine tactile patterns: five taps at different temple locations [\texttt{front, center, back, right, left}] and four sliding directions [\texttt{back-to-front, front-to-back, left-to-right, right-to-left}], as shown in Fig.~\ref{fig:3dview}b. Sliding patterns were rendered at slow (1.5 s) and fast (1 s) speeds, utilizing 13 total patterns created for spatial intuitiveness.
\begin{figure*}[t]
    \vspace*{0.4cm}
    \centering
    \includegraphics[width=0.9\textwidth]{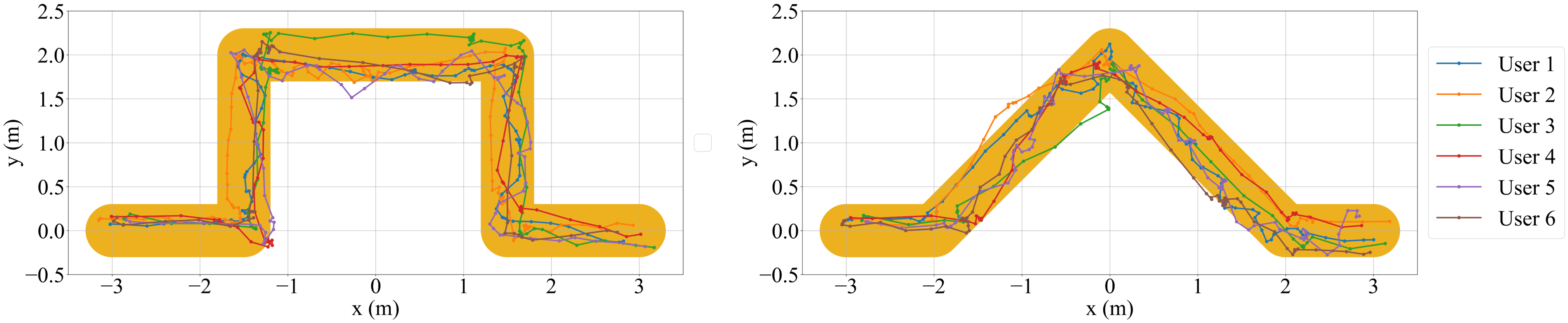} 
    \caption{Participants' trajectories for the two guiding paths. The orange area represents the $\pm 0.3$ m tolerance zone.}
    \Description{Two side-by-side trajectory plots showing participant navigation paths in a 6x6 meter room. Each plot displays multiple overlapping colored lines representing individual participant trajectories, with black dots marking waypoints. The orange-shaded area represents the plus-or-minus 0.3 meter tolerance zone around the intended path. Path 1 on the left contains 6 waypoints forming an irregular path with turns. Path 2 on the right contains 5 waypoints in a similar configuration. Most trajectories stay within or near the orange tolerance zone, demonstrating successful haptic-guided navigation.}
    \label{fig:paths}
\end{figure*}
13 participants (12 male, 1 female, aged 22--35, mean 25.7 $\pm$ 3.8 years) with normal sensorimotor function completed the study after providing informed consent. Following a brief training session with each pattern presented three times, participants identified patterns via a graphical interface in 65 randomized trials (each pattern presented five times).

Recognition accuracy averaged 81\%, ranging from 95\% (back sliding slow) to 65\% (right tap); the full confusion matrix is shown in Table~\ref{table:confusion}. Repeated-measures ANOVA revealed significant differences across patterns ($F(12,156) = 1.9902$, $p = 0.0284$) and between tapping versus sliding overall ($F(1,167) = 10.9792$, $p = 0.0011$), though no differences emerged within each category. Participants frequently reported the ``hanger effect'', where sliding patterns naturally prompted head movements in the indicated direction, suggesting intuitive navigational potential. However, participants were exposed to the system only for short durations, and longer-term effects such as fatigue, pressure discomfort, or habituation to haptic stimuli were not assessed.

\paragraph{Haptic Feedback Navigation Performance}

This evaluation assessed whether users could follow predefined paths using only haptic cues (slide-front, slide-left, slide-right, tap-front) selected based on recognition performance (see Fig. ~\ref{fig:paths}).

6 participants (4 male, 2 female, aged 21--35, mean 25.83 $\pm$ 5.45 years) with normal sensorimotor function provided informed consent and completed calibration and training. In a 6$\times$6 m room with VICON motion capture, participants navigated two paths (6 and 5 waypoints). The system rendered corrective patterns (slide-left/right) when participants deviated beyond $\pm$0.3 m or $\pm$15$^\circ$ orientation thresholds, slide-front for forward motion, and tap-front upon waypoint arrival.

All participants completed both paths successfully (Fig.~\ref{fig:paths}). Average completion times were 2 min 38 s (path 1) and 2 min 9 s (path 2). Participants remained within tolerance 95.73\% (path 1) and 93.34\% (path 2) of the time, with an average of 1.33 and 1.0 boundary crossings per path, respectively. While these results confirm effective haptic guidance interpretation, the 1.25 s pattern execution time limited responsiveness during rapid movements.

\paragraph{LLM and Video-Based Navigation}

\begin{figure}[h]
    \centering
    \includegraphics[width=0.47\textwidth]{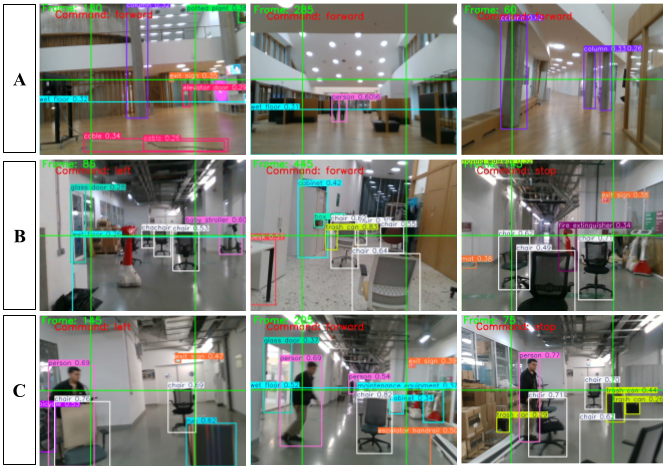} 
    \caption{Example scenarios: (A) no obstacles, (B) static obstacles, (C) dynamic obstacles.}
    \Description{Three photographs showing different navigation scenarios from the user perspective. Image (A) labeled "no obstacles" shows an empty indoor corridor with  walls and clear floor space. Image (B) labeled "static obstacles" shows the same corridor with several chairs positioned as fixed barriers. Image (C) labeled "dynamic obstacles" displays the corridor with people and moveable objects that could change position during navigation. All three images are taken from approximately the same viewpoint to demonstrate increasing environmental complexity.}

    \label{fig:LLMVideo}
    \vspace{-0.4cm}
\end{figure}

This experiment evaluated the integrated vision-language pipeline by comparing system-generated navigation decisions to human judgment across three scenarios: open space, static obstacles, and dynamic obstacles (Fig.~\ref{fig:LLMVideo}). Twenty trials per scenario were conducted.

Decision accuracy was 91.8\% for open environments, 84.6\% with static obstacles, and 81.5\% with dynamic obstacles (e.g., moving chairs). These results demonstrate robust obstacle detection and navigation-cue generation, with performance remaining high even under increased environmental complexity.

\section{Conclusions and Future Work}

This work presents a novel wearable navigation system designed to assist individuals with visual impairments by integrating haptic feedback, generative AI, and real-time object detection. The system features temple-mounted actuators driven by a five-bar linkage mechanism to deliver intuitive tactile cues. An ESP32 CAM camera captures video data, which is processed by a YOLO-World object detection module and a GPT-4o LLM to generate navigation commands. In the current implementation, video streams are processed solely for navigation purposes and are not stored. Additionally, the hanger effect hypothesis is leveraged to enhance the intuitiveness of haptic feedback, guiding users naturally through tactile cues.

Three experimental evaluations validated the system’s effectiveness. The first study assessed the recognition of 13 distinct haptic patterns, including tapping and sliding sensations at varying speeds. Participants achieved an average recognition rate of 81.3\%, with sliding patterns showing the highest accuracy. The second study, conducted in a motion-tracked VICON environment, evaluated the system's ability to guide users along predefined paths. The final evaluation tested the integration of video-based navigation and LLM decision-making under three scenarios: no obstacles, fixed barriers, and moving objects. The system demonstrated high alignment with human decision-making, confirming its reliability in diverse conditions.

While these results highlight the system’s potential, this work represents an early-stage implementation with opportunities for further refinement. Future research will focus on optimizing response times, particularly during rapid user movements, and evaluating performance in more complex environments, such as outdoor settings or varying lighting conditions. Additionally, enhancements to the haptic rendering mechanism and integration of advanced VLMs will improve obstacle detection and navigational precision. 

With continued development, the system has the potential to become a comprehensive navigation aid for visually impaired individuals. By fostering greater independence, confidence, and accessibility, this technology can empower users to navigate complex environments, participate in social activities, and engage more fully in their communities.

\section*{Acknowledgements} 
Research reported in this publication was financially supported by the RSF grant No. 24-41-02039.


%

\balance

\bibliographystyle{ACM-Reference-Format}
\bibliography{bib}


\end{document}

%% file: confussion_matrix.tex
\begin{table*}[]

\centering{
\caption{Confusion Matrix for Actual and Perceived Pattern Recognition.}
\label{table:confusion} 
\setlength{\tabcolsep}{7pt} 
\renewcommand{\arraystretch}{0.8} 

\begin{tabular}{|cccl|ccccccccccccc|}
\hline
\multicolumn{4}{|c|}{}                                                                                                                                  & \multicolumn{13}{c|}{Answers   (Predicted Class)}                                                                                                                                                                                                                                                                                                                                                                                                                                                                                                                                                                                                                                                                                                                                                                                                                                                                                                                                       \\ \cline{5-17} 
\multicolumn{4}{|c|}{}                                                                                                                                  & \multicolumn{5}{c|}{}                                                                                                                                                                                                                                                                                                                                                                & \multicolumn{8}{c|}{Slide}                                                                                                                                                                                                                                                                                                                                                                                                                                                                                                                                                                       \\ \cline{10-17} 
\multicolumn{4}{|c|}{}                                                                                                                                  & \multicolumn{5}{c|}{\multirow{-2}{*}{Tap}}                                                                                                                                                                                                                                                                                                                                           & \multicolumn{4}{c|}{Fast}                                                                                                                                                                                                                                                                                 & \multicolumn{4}{c|}{Slow}                                                                                                                                                                                                                                                            \\ \cline{5-17} 
\multicolumn{4}{|c|}{\multirow{-4}{*}{\%}}                                                                                                              & \multicolumn{1}{c|}{front}                                               & \multicolumn{1}{c|}{center}                                              & \multicolumn{1}{c|}{back}                                                & \multicolumn{1}{c|}{left}                                                & \multicolumn{1}{c|}{right}                                               & \multicolumn{1}{c|}{front}                                               & \multicolumn{1}{c|}{back}                                                & \multicolumn{1}{c|}{right}                                               & \multicolumn{1}{c|}{left}                                                & \multicolumn{1}{c|}{front}                                               & \multicolumn{1}{c|}{back}                                                & \multicolumn{1}{c|}{right}                                               & left                                                \\ \hline
\multicolumn{1}{|c|}{}                            & \multicolumn{2}{c|}{}                                                                      & front  & \multicolumn{1}{c|}{\cellcolor[HTML]{0B3040}{\color[HTML]{FFFFFF} 0.85}} & \multicolumn{1}{c|}{\cellcolor[HTML]{E0E5E7}0.11}                        & \multicolumn{1}{c|}{\cellcolor[HTML]{FBFCFC}0.02}                        & \multicolumn{1}{c|}{\cellcolor[HTML]{F7F8F9}0.03}                        & \multicolumn{1}{c|}{-}                        & \multicolumn{1}{c|}{-}                        & \multicolumn{1}{c|}{-}                        & \multicolumn{1}{c|}{-}                        & \multicolumn{1}{c|}{-}                        & \multicolumn{1}{c|}{-}                        & \multicolumn{1}{c|}{-}                        & \multicolumn{1}{c|}{-}                        & -                        \\ \hhline{~~~--------------}
\multicolumn{1}{|c|}{}                            & \multicolumn{2}{c|}{}                                                                      & center & \multicolumn{1}{c|}{\cellcolor[HTML]{EEF0F2}0.06}                        & \multicolumn{1}{c|}{\cellcolor[HTML]{193C4B}{\color[HTML]{FFFFFF} 0.80}} & \multicolumn{1}{c|}{\cellcolor[HTML]{F2F4F5}0.05}                        & \multicolumn{1}{c|}{\cellcolor[HTML]{E9EDEE}0.08}                        & \multicolumn{1}{c|}{\cellcolor[HTML]{FBFCFC}0.02}                        & \multicolumn{1}{c|}{-}                        & \multicolumn{1}{c|}{-}                        & \multicolumn{1}{c|}{-}                        & \multicolumn{1}{c|}{-}                        & \multicolumn{1}{c|}{-}                        & \multicolumn{1}{c|}{-}                        & \multicolumn{1}{c|}{-}                        & -                        \\ \hhline{~~~--------------}
\multicolumn{1}{|c|}{}                            & \multicolumn{2}{c|}{}                                                                      & back   & \multicolumn{1}{c|}{-}                        & \multicolumn{1}{c|}{\cellcolor[HTML]{C1CBCF}0.22}                        & \multicolumn{1}{c|}{\cellcolor[HTML]{335260}{\color[HTML]{FFFFFF} 0.71}} & \multicolumn{1}{c|}{\cellcolor[HTML]{FBFCFC}0.02}                        & \multicolumn{1}{c|}{\cellcolor[HTML]{F2F4F5}0.05}                        & \multicolumn{1}{c|}{-}                        & \multicolumn{1}{c|}{\cellcolor[HTML]{FBFCFC}0.02}                        & \multicolumn{1}{c|}{-}                        & \multicolumn{1}{c|}{-}                        & \multicolumn{1}{c|}{-}                        & \multicolumn{1}{c|}{-}                        & \multicolumn{1}{c|}{-}                        & -                        \\ \hhline{~~~--------------} 
\multicolumn{1}{|c|}{}                            & \multicolumn{2}{c|}{}                                                                      & left   & \multicolumn{1}{c|}{\cellcolor[HTML]{E9EDEE}0.08}                        & \multicolumn{1}{c|}{\cellcolor[HTML]{DCE1E4}0.12}                        & \multicolumn{1}{c|}{\cellcolor[HTML]{F2F4F5}0.05}                        & \multicolumn{1}{c|}{\cellcolor[HTML]{2F4F5C}{\color[HTML]{FFFFFF} 0.72}} & \multicolumn{1}{c|}{\cellcolor[HTML]{F7F8F9}0.03}                        & \multicolumn{1}{c|}{-}                        & \multicolumn{1}{c|}{-}                        & \multicolumn{1}{c|}{-}                        & \multicolumn{1}{c|}{-}                        & \multicolumn{1}{c|}{-}                        & \multicolumn{1}{c|}{-}                        & \multicolumn{1}{c|}{-}                        & -                        \\ \hhline{~~~--------------} 
\multicolumn{1}{|c|}{}                            & \multicolumn{2}{c|}{\multirow{-5}{*}{{\rotatebox{90}{Tap}}}}                                                 & right  & \multicolumn{1}{c|}{\cellcolor[HTML]{F7F8F9}0.03}                        & \multicolumn{1}{c|}{\cellcolor[HTML]{C1CBCF}0.22}                        & \multicolumn{1}{c|}{\cellcolor[HTML]{E5E9EB}0.09}                        & \multicolumn{1}{c|}{-}                        & \multicolumn{1}{c|}{\cellcolor[HTML]{45616E}{\color[HTML]{FFFFFF} 0.65}} & \multicolumn{1}{c|}{-}                        & \multicolumn{1}{c|}{-}                        & \multicolumn{1}{c|}{-}                        & \multicolumn{1}{c|}{}                                                    & \multicolumn{1}{c|}{-}                        & \multicolumn{1}{c|}{-}                        & \multicolumn{1}{c|}{-}                        & -                        \\ \hhline{~----------------} 
\multicolumn{1}{|c|}{}                            & \multicolumn{1}{c|}{}                        & \multicolumn{1}{c|}{}                       & front  & \multicolumn{1}{c|}{-}                        & \multicolumn{1}{c|}{-}                        & \multicolumn{1}{c|}{-}                        & \multicolumn{1}{c|}{-}                        & \multicolumn{1}{c|}{-}                        & \multicolumn{1}{c|}{\cellcolor[HTML]{193C4B}{\color[HTML]{FFFFFF} 0.80}} & \multicolumn{1}{c|}{-}                        & \multicolumn{1}{c|}{\cellcolor[HTML]{E5E9EB}0.09}                        & \multicolumn{1}{c|}{\cellcolor[HTML]{FBFCFC}0.02}                        & \multicolumn{1}{c|}{\cellcolor[HTML]{E5E9EB}0.09}                        & \multicolumn{1}{c|}{-}                        & \multicolumn{1}{c|}{-}                        & -                        \\ \hhline{~~~--------------} 
\multicolumn{1}{|c|}{}                            & \multicolumn{1}{c|}{}                        & \multicolumn{1}{c|}{}                       & back   & \multicolumn{1}{c|}{-}                        & \multicolumn{1}{c|}{-}                        & \multicolumn{1}{c|}{-}                        & \multicolumn{1}{c|}{-}                        & \multicolumn{1}{c|}{-}                        & \multicolumn{1}{c|}{-}                        & \multicolumn{1}{c|}{\cellcolor[HTML]{0B3040}{\color[HTML]{FFFFFF} 0.83}} & \multicolumn{1}{c|}{\cellcolor[HTML]{F6F8F8}0.03}                        & \multicolumn{1}{c|}{\cellcolor[HTML]{F6F8F8}0.03}                        & \multicolumn{1}{c|}{-}                        & \multicolumn{1}{c|}{\cellcolor[HTML]{E4E8EA}0.09}                        & \multicolumn{1}{c|}{\cellcolor[HTML]{FBFCFC}0.02}                        & -                        \\ \hhline{~~~--------------} 
\multicolumn{1}{|c|}{}                            & \multicolumn{1}{c|}{}                        & \multicolumn{1}{c|}{}                       & right  & \multicolumn{1}{c|}{-}                        & \multicolumn{1}{c|}{-}                        & \multicolumn{1}{c|}{\cellcolor[HTML]{FBFCFC}0.02}                        & \multicolumn{1}{c|}{-}                        & \multicolumn{1}{c|}{-}                        & \multicolumn{1}{c|}{-}                        & \multicolumn{1}{c|}{\cellcolor[HTML]{FBFCFC}0.02}                        & \multicolumn{1}{c|}{\cellcolor[HTML]{0B3040}{\color[HTML]{FFFFFF} 0.89}} & \multicolumn{1}{c|}{-}                        & \multicolumn{1}{c|}{-}                        & \multicolumn{1}{c|}{-}                        & \multicolumn{1}{c|}{\cellcolor[HTML]{EAEEEF}0.08}                        & -                        \\ \hhline{~~~--------------} 
\multicolumn{1}{|c|}{}                            & \multicolumn{1}{c|}{}                        & \multicolumn{1}{c|}{\multirow{-4}{*}{{\rotatebox{90}{Fast}}}} & left   & \multicolumn{1}{c|}{-}                        & \multicolumn{1}{c|}{\cellcolor[HTML]{FBFCFC}0.02}                        & \multicolumn{1}{c|}{-}                        & \multicolumn{1}{c|}{-}                        & \multicolumn{1}{c|}{-}                        & \multicolumn{1}{c|}{-}                        & \multicolumn{1}{c|}{-}                        & \multicolumn{1}{c|}{\cellcolor[HTML]{FBFCFC}0.02}                        & \multicolumn{1}{c|}{\cellcolor[HTML]{0B3040}{\color[HTML]{FFFFFF} 0.89}} & \multicolumn{1}{c|}{-}                        & \multicolumn{1}{c|}{-}                        & \multicolumn{1}{c|}{\cellcolor[HTML]{FBFCFC}0.02}                        & \cellcolor[HTML]{EFF1F2}0.06                        \\ \hhline{~~---------------}
\multicolumn{1}{|c|}{}                            & \multicolumn{1}{c|}{}                        & \multicolumn{1}{c|}{}                       & front  & \multicolumn{1}{c|}{-}                        & \multicolumn{1}{c|}{-}                        & \multicolumn{1}{c|}{\cellcolor[HTML]{FBFCFC}0.02}                        & \multicolumn{1}{c|}{\cellcolor[HTML]{FBFCFC}0.02}                        & \multicolumn{1}{c|}{-}                        & \multicolumn{1}{c|}{\cellcolor[HTML]{DFE4E6}0.11}                        & \multicolumn{1}{c|}{-}                        & \multicolumn{1}{c|}{-}                        & \multicolumn{1}{c|}{-}                        & \multicolumn{1}{c|}{\cellcolor[HTML]{0B3040}{\color[HTML]{FFFFFF} 0.82}} & \multicolumn{1}{c|}{\cellcolor[HTML]{F2F4F5}0.05}                        & \multicolumn{1}{c|}{-}                        & -                        \\ \hhline{~~~--------------} 
\multicolumn{1}{|c|}{}                            & \multicolumn{1}{c|}{}                        & \multicolumn{1}{c|}{}                       & back   & \multicolumn{1}{c|}{-}                        & \multicolumn{1}{c|}{-}                        & \multicolumn{1}{c|}{-}                        & \multicolumn{1}{c|}{-}                        & \multicolumn{1}{c|}{-}                        & \multicolumn{1}{c|}{-}                        & \multicolumn{1}{c|}{\cellcolor[HTML]{F8F9F9}0.03}                        & \multicolumn{1}{c|}{-}                        & \multicolumn{1}{c|}{-}                        & \multicolumn{1}{c|}{-}                        & \multicolumn{1}{c|}{\cellcolor[HTML]{0B3040}{\color[HTML]{FFFFFF} 0.95}} & \multicolumn{1}{c|}{\cellcolor[HTML]{FCFCFC}0.02}                        & -                        \\ \hhline{~~~--------------} 
\multicolumn{1}{|c|}{}                            & \multicolumn{1}{c|}{}                        & \multicolumn{1}{c|}{}                       & right  & \multicolumn{1}{c|}{-}                        & \multicolumn{1}{c|}{-}                        & \multicolumn{1}{c|}{-}                        & \multicolumn{1}{c|}{-}                        & \multicolumn{1}{c|}{-}                        & \multicolumn{1}{c|}{-}                        & \multicolumn{1}{c|}{-}                        & \multicolumn{1}{c|}{\cellcolor[HTML]{E9ECEE}0.08}                        & \multicolumn{1}{c|}{\cellcolor[HTML]{F6F8F8}0.03}                        & \multicolumn{1}{c|}{-}                        & \multicolumn{1}{c|}{\cellcolor[HTML]{FBFCFC}0.02}                        & \multicolumn{1}{c|}{\cellcolor[HTML]{0B3040}{\color[HTML]{FFFFFF} 0.83}} & \cellcolor[HTML]{F2F4F5}0.05                        \\ \hhline{~~~--------------} 
\multicolumn{1}{|c|}{\multirow{-13}{*}{{\rotatebox{90}{Patterns}}}} & \multicolumn{1}{c|}{\multirow{-8}{*}{{\rotatebox{90}{Slide}}}} & \multicolumn{1}{c|}{\multirow{-4}{*}{{\rotatebox{90}{Slow}}}} & left   & \multicolumn{1}{c|}{-}                        & \multicolumn{1}{c|}{-}                        & \multicolumn{1}{c|}{-}                        & \multicolumn{1}{c|}{-}                        & \multicolumn{1}{c|}{-}                        & \multicolumn{1}{c|}{\cellcolor[HTML]{FBFCFC}0.02}                        & \multicolumn{1}{c|}{\cellcolor[HTML]{FBFCFC}0.02}                        & \multicolumn{1}{c|}{\cellcolor[HTML]{FBFCFC}0.02}                        & \multicolumn{1}{c|}{\cellcolor[HTML]{E9EDEE}0.08}                        & \multicolumn{1}{c|}{-}                        & \multicolumn{1}{c|}{\cellcolor[HTML]{FBFCFC}0.02}                        & \multicolumn{1}{c|}{\cellcolor[HTML]{F7F8F9}0.03}                        & \cellcolor[HTML]{103444}{\color[HTML]{FFFFFF} 0.83} \\ \hline
\end{tabular}}
\vspace{-0.3cm}
\end{table*}